# Control of electronic conduction at an oxide heterointerface using surface polar adsorbates

Yanwu Xie[1], Yasuyuki Hikita[1], Christopher Bell[1,3] and Harold Y. Hwang[1,2,3]

[1] *Department of Advanced Materials Science, University of Tokyo*

*Kashiwa, Chiba 277-8561, Japan*

[2] *Japan Science and Technology Agency, Kawaguchi, Saitama 332-0012, Japan*

[3] *Department of Applied Physics and Stanford Institute for Materials and Energy Science,*

*Stanford University, Stanford, California 94305, USA*

**The transfer of electrons between a solid surface and adsorbed atomic or molecular species is fundamental in natural and synthetic processes, being at the heart of most catalytic reactions and many sensors. In special cases, metallic conduction can be induced at the surface of, for example, Si-terminated SiC[1], or mixed-terminated ZnO[2], in the presence of a hydrogen adlayer. Generally, only the surface atoms are significantly affected by adsorbates. However, remotely changing electronic states far from the adsorbed layer is possible if these states are electrostatically coupled to the surface. Here we show that the surface adsorption of common solvents such as acetone, ethanol, and water can induce a large change (factor of three) in the conductivity at the buried interface between $SrTiO_3$ substrates and $LaAlO_3$ thin films[3-8]. This phenomenon is observed only for polar solvents. Our result provides experimental evidence that adsorbates at the $LaAlO_3$ surface induce accumulation of electrons at the $LaAlO_3/SrTiO_3$ interface, suggesting a general polarization-facilitated electronic transfer mechanism, which can be used for sensor applications.**

The intense interest in the $LaAlO_3/SrTiO_3$ interface[3-20] has led to several recent experimental observations which suggest a close relationship between the interface and the $LaAlO_3$ surface. These include the use of conducting atomic force microscopy probes to toggle a metal insulator transition[10] through the writing of surface charge[11], and the use of capping layers of $SrTiO_3$[12] or $SrCuO_2$[13] to alter the interface conductivity. The origin of these effects, and indeed the interface conductivity itself, remains in debate[4, 14-20]. The transfer of electrons from the surface to the interface, to reconcile the polar discontinuity between the neutral {100} layers in $SrTiO_3$ and the charged layers in {100} $LaAlO_3$, is one model that naturally connects the electronic states of the interface and $LaAlO_3$ surface[4,16-18]. However, despite these concepts and recent experimental progress, the effect of the surface adsorbate has not been investigated. This issue is addressed in the present work. We have found that the exposure of LaAlO$_3$/SrTiO$_3$ samples to a polar solvent can increase the sheet carrier density, $n_{2d}$, by more than $2\times10^{13}$ cm$^{-2}$, representing a change of the same order as the total charge density typical in this system[4-8]. Compared with the strong perturbations associated with other surface processes, *i.e.*, the extremely strong local electric field produced by the conducting atomic force microscopy probes[10,11], or the structural variation by introducing capping layers grown at high temperatures[12,13], the changes associated with room temperature treatment using these solvents would naively be expected to be small. However, such processes result in a surprisingly large modulation of $n_{2d}$, revealing a dramatic surface-interface coupling.

The fabrication of the conducting LaAlO$_3$/SrTiO$_3$ interfaces and the surface adsorption process (SAP) are described in the Methods. As shown in Fig. 1a, labels 1 & 2, a SAP using acetone increased $n_{2d}$ from $\sim1\times10^{13}$ cm$^{-2}$ to more than $3\times10^{13}$ cm$^{-2}$, over a wide temperature range (2 K $\leq T \leq$ 300 K). This remarkable modulation suggests that the acetone molecules have been adsorbed on the LaAlO$_3$ surface and in turn change the electronic states at the buried interface. Heating the sample at 380 K in a moderate vacuum (< $10^3$ Pa, with a helium background) for several hours produced negligible conduction change (not shown). However, the original sample state was recovered after heating at an elevated temperature of 653 K, evidenced by the decrease in $n_{2d}$ (Fig. 1a, label 3) and the subsequent increase after another SAP using water (Fig.1a, label 4). These facts indicate that the surface adsorption is reversible but strong, not a physical adsorption driven by the relatively weak van der Waals force, and thus likely involves electron transfer.

Accompanying the $n_{2d}$ increase, a striking reduction in the Hall mobility, $\mu_H$, from 3,000 cm$^2$V$^{-1}$s$^{-1}$ to 600 cm$^2$V$^{-1}$s$^{-1}$ ($T$ = 2 K) is observed after the acetone SAP (Fig. 1b). $\mu_H$ decreases quickly with increasing temperature, with a relatively small room temperature value of $\mu_H \sim 6$ cm$^2$V$^{-1}$s$^{-1}$ due to the dominance of phonon scattering. At this temperature $\mu_H$ also has a very weak dependence on $n_{2d}$, thus the conductivity $1/R_{sheet}$ is a reasonable and convenient index of $n_{2d}$ (Figs. 1a & c), and we will use it hereafter to characterize the effect of the SAP process on $n_{2d}$ for a variety of common solvents. To study the repeatability of the adsorption / desorption processes, we repeatedly performed water SAP on a LaAlO$_3$/SrTiO$_3$ sample followed by a heating step at 573 K in an oxygen flow, to reset the system (Fig. 1d). The sample switches reproducibly between a high-conductivity state after SAP and a low-conductivity state after heating. We note that this behavior, together with the fast SAP response (less than seconds, see Methods), may find use in sensor applications.

To explore the origin of the SAP induced $n_{2d}$, we have studied the effect of SAP using a variety of solvents. These solvents can be classified into three categories: non-polar, polar aprotic (no dissociable H$^+$), and polar protic (dissociable H$^+$). The results are summarized in Fig. 2. Clearly, SAP using non-polar solvents produces no significant modulation in $1/R_{sheet}$ and by extension, $n_{2d}$. SAP using a relatively weak polar solvent, such as ethyl acetate, produces a slight increase in $n_{2d}$, while using more polar solvents produces a much larger increase, independent of whether the solvents are aprotic or protic. In Fig. 2 we also show that $\Delta(1/R_{Sheet})$ is well scaled with the molecular dipole moment divided by the molecular volume. These results suggest that the polar nature of solvents plays a key role in determining the magnitude of the observed changes for the different solvents. In the context of an electrostatic mechanism, the scaling of the change with the dipole moment density would be reasonable, since the strength of the effect should be related to not only the molecular dipole moment, but also the areal density of the molecules on the surface.

Additional insights into the origin of the SAP induced $n_{2d}$ can be made by comparing the effect of SAP with the effect of adsorption from a solvent vapor. It is well known that a thin layer of water coats all hydrophilic surfaces under ambient conditions[21]. While $La_2O_3$ is highly hygroscopic[22], $Al_2O_3$ adsorbs one monolayer of water when the relative humidity (RH) is ~35%, and more than five monolayers for RH > 70%[23,24]. Thus, although these studies cannot be trivially extrapolated to the current heterostructures, we expect a thin layer of water with full coverage will coat the $AlO_2$-terminated $LaAlO_3$ surface when the $LaAlO_3/SrTiO_3$ sample is exposed to RH > 70% air at room temperature. Such a surface layer has also been inferred from charge writing experiments[25]. However, Fig. 3a clearly shows that the conductivity of $LaAlO_3/SrTiO_3$ is still sensitive to water SAP using liquid even after exposing it to water saturated atmosphere for several hours. A similar result was found when acetone was used (Fig. 3b). From these results we can conclude that the thickness of adsorbate layer also plays an important role in the SAP effect, since the SAP with liquid is expected to produce a thicker adsorbed solvent film compared to the vapor experiments.

Next we discuss the possible mechanisms that may explain these data. We exclude electrostatic attraction since the polar adsorbates themselves are charge-neutral as a whole and will not change the electrostatic boundary condition of $LaAlO_3/SrTiO_3$ without the transfer of electrons[26]. Oxygen vacancies[14], interdiffusion[15], and an electronic reconstruction[4,16-18] have all been suggested to be the origin of the interfacial conductivity in the $LaAlO_3/SrTiO_3$ system. As far as the SAP experiments, we exclude the first two of these since we assume that a SAP at room temperature does not cause significant atomic structure changes at the interface, and we focus on the electronic reconstruction mechanism.

As discussed in several papers[17,18], and illustrated in Fig. 4a, for an idealized LaAlO$_3$ surface free from surface states, the valence band of LaAlO$_3$, $E_V^{LAO}$, is the electron source for the interface electrons. Here we assume no conductivity in the LaAlO$_3$ film itself, and for simplicity do not consider other processes that do not affect the electrostatic state of the LaAlO$_3$. An uncompensated potential, $V_{Uncom}$, across the LaAlO$_3$ film of the order of the band gap of SrTiO$_3$, $E_g^{STO}$, is needed to align the Fermi level, $E_F$, over the whole structure. A simple electrostatic consideration shows that $V_{Uncom} = \frac{e\left(\frac{\sigma_0}{2} - \sigma_{Inter}\right)}{\varepsilon_{LAO}} d_{LAO}$, where $\sigma_0$ is the charge density of one atomic layer, $\sigma_{Inter}$ is the total sheet electron density at interface, $e$ is the unit charge, and $\varepsilon_{LAO}$ is the dielectric constant of LaAlO$_3$. In reality $\sigma_{Inter}$ is larger than $n_{2d}$ due to the trapping of some electrons at the interface[17]. Immediately, we find that

$$\sigma_{Inter} = \frac{\sigma_0}{2} - \frac{\varepsilon_{LAO}}{e\, d_{LAO}} V_{Uncom} \quad (1).$$

This simple relationship predicts the same thickness dependence of $\sigma_{Inter}$ as several more detailed theoretical calculations[18,27], noting that $V_{Uncom}$ is fixed for an ideal surface. In the work presented here, $d_{LAO}$ is fixed, and the SAP induced change in $n_{2d}$ is thus associated with a reduction in $V_{Uncom}$. Parenthetically, two recent theoretical papers have suggested that the dissociation of hydrogen[19] and redox reactions[20] on the LaAlO$_3$ surface as possible electron sources. In both cases $V_{Uncom}$ is reduced compared with that of the idealized surface because the energy level of the alternative electron sources is higher than $E_V^{LAO}$.

With the above picture in mind, we can interpret our results as follows. The strong polarization in the LaAlO$_3$ layer aligns to a degree the polar adsorbates nearest the surface, with a decaying trend as we move away into the adsorbate layer. As shown schematically in Fig. 4b, this leads to an electrostatic potential across the adsorbate layer, $V_{ad}$, quite similar to the built-in potential in the polar LaAlO$_3$. As a result, $V_{Uncom}$ across the LaAlO$_3$ is effectively decreased, and thus $\sigma_{Inter}$ increases according to equation (1). This simple mechanism consistently explains the observed SAP induced accumulation of electrons at the LaAlO$_3$/SrTiO$_3$ interface.

The observed close coupling between polar surface adsorbates and interface conductivity provides new insights into the origin of electron gas at the LaAlO$_3$/SrTiO$_3$ interface, and demonstrates a new tuning parameter in controlling this exotic electron gas. These data also suggest that many of the conflicting (and often contradictory) studies of this popular system can be explained by this observation – *i.e.* whether measurements are made *in-situ*[28], or after various surface exposures[29, 30], the sensitive electronic structure of the interface is dramatically altered.

**Methods**

**Sample fabrication.** Using optical lithography and lift-off, a six contact Hall bar (central bar length 50 μm, width 10 μm) with amorphous AlO$_x$ as a hard mask was patterned onto the atomically flat TiO$_2$-terminated SrTiO$_3$ (100) substrates. The LaAlO$_3$ thin films were grown on the patterned substrates by pulsed laser deposition with the growth monitored by *in-situ* reflection high-energy electron diffraction. The conducting interfaces are confined within the Hall bar region. For the samples shown in this Letter, the thickness was fixed at 10 unit cells, and the growth was at 923 K in an O$_2$ pressure of $1.33 \times 10^{-3}$ Pa, after a pre annealing at 1223 K in an $O_2$ pressure of $6.67 \times 10^{-4}$ Pa for half an hour. After deposition there followed a post annealing step at 873 K in an $O_2$ pressure of $4 \times 10^4$ Pa for one hour. During deposition the laser repetition was 1 Hz and the laser energy density was 0.6 Jcm$^{-2}$. The same phenomenology is found for samples grown in a variety of temperatures, oxygen pressures and LaAlO$_3$ thicknesses.

**Electrical contact and measurement.** The conducting interface was contacted by ultrasonic bonding with Al wire. The temperature-dependent sheet resistance, sheet carrier density and mobility were deduced from standard transport measurements using the patterned Hall bar. All transport measurements were carried out using a standard four-probe method. To test for the possibility of parallel conductivity through the adsorbate layer itself, contacts were made directly to the surface using silver epoxy. No significant surface conduction was detected both before and after SAP, clearly demonstrating that the conduction is at all times dominated by the LaAlO$_3$/SrTiO$_3$ interface.

**Surface adsorption process.** SAP was achieved by placing a drop of the liquid solvent on the sample surface at room temperature and blowing off all visible solvent within less than ten seconds using dry nitrogen gas. Contacts were either left on the samples during the SAP, or reconnected afterwards, no significant difference in the conduction was observed by using either method.

**Acknowledgements**

We thank N. Ogawa for useful discussions and K. Nishio, R. Takahashi, and M. Lippmaa for technical assistance. Y.W.X. acknowledges funding from the Japan Society for the Promotion of Science (JSPS). C.B. and H.Y.H. acknowledge support by the Department of Energy, Office of Basic Energy Sciences, under Contract No. DE-AC02-76SF00515.

**Author contributions**

Y.W.X. performed sample fabrication, measurements, and data analysis. Y.H., C.B. and H.Y.H. assisted with the planning, measurements, and analysis.

**Additional Information**

The authors declare no competing financial interests. Correspondence and requests for materials should be addressed to Y.W.X. (xie@ams.k.u-tokyo.ac.jp).

**Figure legends**

**Figure 1 | Transport characterization.** Temperature dependence of **a**, Sheet carrier density, $n_{2d}$, **b**, Hall mobility, $\mu_H$, and **c**, Sheet resistance, $R_{Sheet}$, in different states of a 10 unit cell LaAlO$_3$/SrTiO$_3$ sample. The measurement sequence follows the labeled order. "As grown" corresponds to the sample before experiencing any SAP. "Acetone" ("water") corresponds to the sample after an acetone (water) SAP process. "Heating" corresponds to $T = 663$ K in an O$_2$ pressure of $1.33 \times 10^{-2}$ Pa for 5 hours. **d**, Room-temperature conductivity, $1/R_{sheet}$, of a LaAlO$_3$/SrTiO$_3$ sample which was repeatedly processed by water SAP and recovered by heating at 573 K in an oxygen flow for 3 hours.

**Figure 2 | Effect of SAP using different solvents.** The difference in $1/R_{sheet}$ after and before SAP, $\Delta(1/R_{sheet})$, reflects the change in $n_{2d}$ at room temperature. A clear increase in $\Delta(1/R_{sheet})$ is observed after SAP using polar solvents, independent of their aprotic or protic character. $\Delta(1/R_{sheet})$ is reasonably scaled with the molecular dipole moment (*D*) of each solvent divided by the corresponding molecular volume ($V_m$) (crosses). The line is a guide to the eye. In each category the solvents are ordered from left to right by increasing polarizability. At least two samples were used for each solvent and all samples were in the "as grown" state before SAP.

**Figure 3 | Comparison between the effect of SAP and the effect of adsorption of solvent vapor.** Conductivity versus time plots for samples exposed to **a**, Water, **b**, Acetone vapours near the saturation point, followed by SAP. Note that surface adsorption of, *e.g.* water, would have already taken place when the sample was kept in normal air. The exposure to the nearly saturated water (acetone) vapour for many hours should achieve a full and thicker coverage of adsorbates on the LaAlO$_3$ surface. The measurements were performed at room temperature.

**Figure 4 | Sketches of band diagrams and electron transfer mechanisms. a**, Idealized surface. Electrons transfer from $E_V^{LAO}$ of surface LaAlO$_3$ to the conduction band of SrTiO$_3$, $E_C^{STO}$, near the interface. $eV_{Uncom}$ roughly equals $E_g^{STO}$. **b**, Surface with aligned polar adsorbates. The built-in potential across the adsorbates, $V_{ad}$, effectively reduces $V_{Uncom}$ across the LaAlO$_3$ layer. In both **a** and **b**, bulk stoichiometry and valences are shown for simplicity.

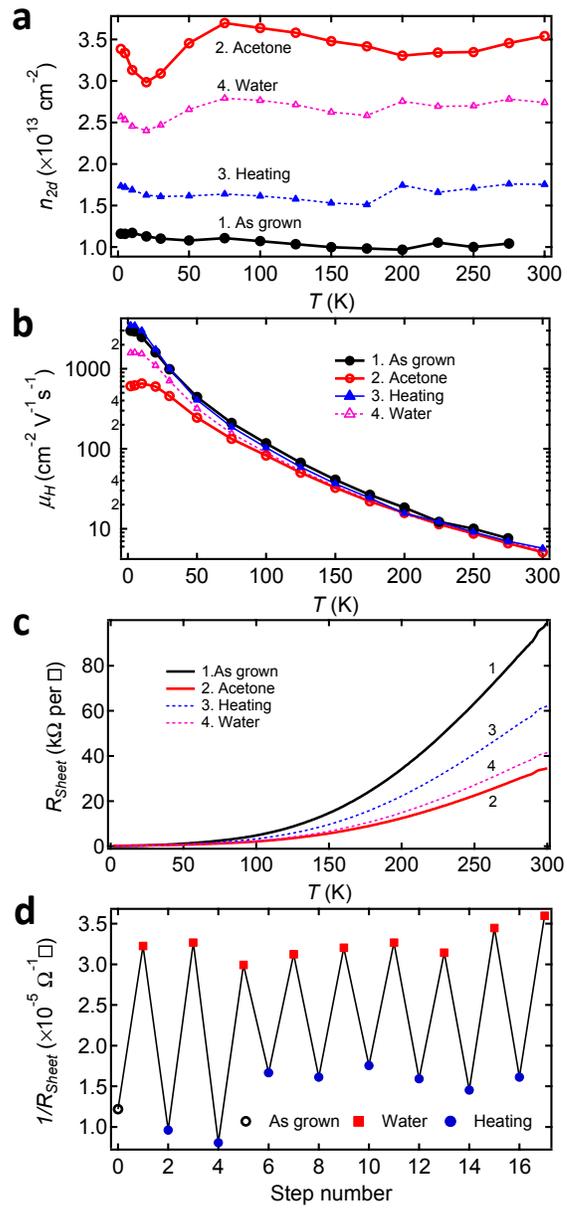

Figure 1 Y. W. Xie *et al.*

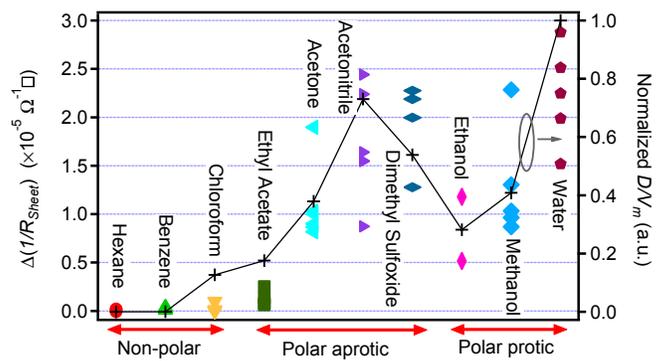

Figure 2 Y. W. Xie *et al.*

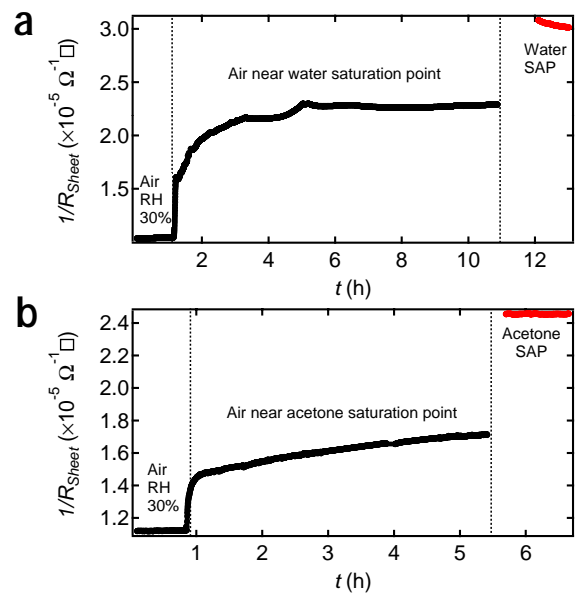

Figure 3 Y. W. Xie *et al.*

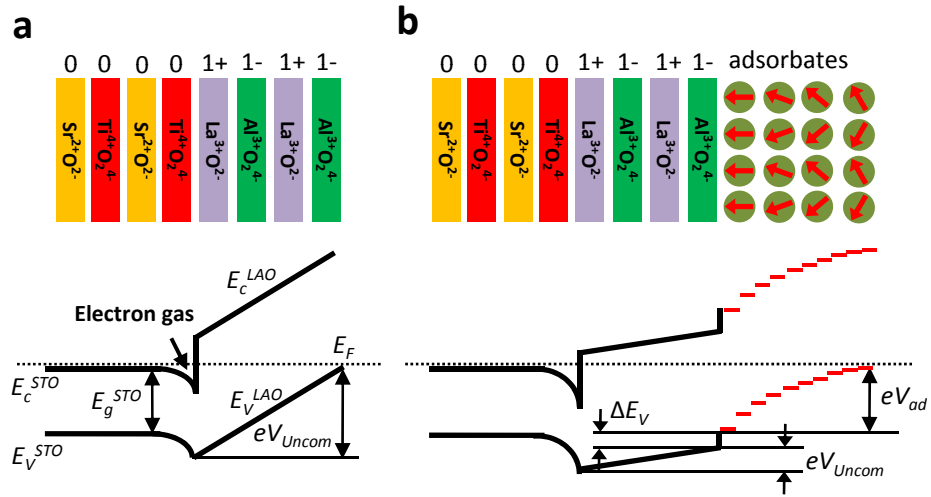

Figure 4 Y. W. Xie *et al.*